\documentclass[graybox]{svmult}

\usepackage{mathptmx}       
\usepackage{helvet}         
\usepackage{courier}        
\usepackage{type1cm}        
\usepackage{makeidx}         
\usepackage{graphicx}        
\usepackage{multicol}        
\usepackage[bottom]{footmisc}

\usepackage{natbib}

\makeindex             

\newcommand\lesim{\lower.5ex\hbox{$\; \buildrel < \over \sim \;$}}
\newcommand\gesim{\lower.5ex\hbox{$\; \buildrel > \over \sim \;$}}
\def\simgt{\lower.5ex\hbox{$\; \buildrel > \over \sim \;$}}
\def\simlt{\lower.5ex\hbox{$\; \buildrel < \over \sim \;$}}
\newcommand\HI{H{\small I}}

\def\subr #1{_{{\rm #1}}}

\def\exp{{\rm \;exp\;}}
\def\gt{{\rm \;>\;}}

\def\D{\partial}
\def\gsim{\lower.5ex\hbox{$\; \buildrel > \over \sim \;$}}
\def\lsim{\lower.5ex\hbox{$\; \buildrel < \over \sim \;$}}
\def\O{\over}

\newcommand\aap{{\it Astron.~Astrophys.}}%
\newcommand\aaps{{\it Astron.~Astrophys.~Suppl.}}%
\newcommand\aspcs{{\it Astron.~Soc.~Pac.~Conf.~Ser.}}%
\newcommand\aj{{\it A.~J.}}%
\newcommand\apj{{\it Astroph.~J.}}%
\newcommand\mnras{{\it Mon.~Not.~R.A.S.}}%
\newcommand\na{{\it New~Astron.}}%

\begin{document}

\title*{Structure, mass and stability of galactic disks}
\titlerunning{Galactic disks}
\author{{\large P.C. van der Kruit}\\
Kapteyn Astronomical Institute, University of Groningen,\\
P.O.Box 800, 9700AV Groningen, the Netherlands,\\ 
\email{vdkruit@astro.rug.nl}}
%
%
\maketitle

\abstract{In this review I concentrate on three areas related to structure
of disks in spiral galaxies. First I will review the work on structure, kinematics and
dynamics of stellar disks. Next I will review the progress in the area of flaring of \HI\ 
layers. These subjects are relevant for the presence of dark matter and
lead to the conclusion that disk are in general not `maximal',
have lower $M/L$ ratios than previously suspected and are locally stable w.r.t. Toomre's
$Q$ criterion for local stability. I will end with a few words on `truncations' in 
stellar disks.}


\section{Stellar disks}
\label{sec:1}

The structure and general 
properties of stellar disks have been reviewed in
some detail by Ken and myself at previous honorary symposia \citep[e.g.][]{vdk02,kcf07}. 
The radial distribution of surface brightness
can be approximated by an exponential \citep{kcf70} and the vertical distribution with an
isothermal sheet \citep{vdks81a}
with a scaleheight that is independent of galactocentric
distance. A more general form  is \citep{vdk88}
\begin{equation}
L(R,z) = L(0,0) {\rm \;e}^{-R/h} \;{\rm sech} ^{2/n} \left( {{n z} \O
{2 h_{\rm z}}} \right),
\label{eqn:vdksdisk}
\end{equation}
This ranges from the isothermal distribution ($n = 1$) to the exponential
function ($n = \infty $) and allows for more realistic 
stellar distribution that are not exactly isothermal in $z$. 
From actual fits in $I$ and $K'$ \citet{dgpvdk97}
found
\begin{equation}
2/n = 0.54 \pm 0.20.
\end{equation}
A detailed study by \citet{dgp97} has shown that the constancy of the vertical
scaleparameter $h_{\rm z}$ is very accurate in late-type spiral disks, but
in early-types it may increase, in an extreme case by  as much as 50\%\ per scalelength $h$.

The origin of the exponential nature of stellar disks is still uncertain. 
\citet{kcf70, kcf75} already pointed out that the
distribution of angular momentum in a self-gravitating exponential disk
resembles that of the uniform, uniformly rotating sphere \citep{mes63}.
This also holds for an exponential density distribution with
a flat rotation curve \citep{gun82, vdk87}, so that a model with
collapse with detailed conservation of angular momentum
\citep[see also][]{fe80} would give a
natural explanation for the exponential nature of disks (and maybe
their truncations; see below). However, bars or other non-axisymmetric
structures may induce severe redistribution of angular momentum;
the work on the effects of nonaxisymmetric
instabilities on the secular evolution of disks and their structural
parameters by \citet{dmcmwq06} shows the potential of such approaches.

The distribution of the scale parameters is most easily studied in edge-on
galaxies. Following on from the studies by \citet{vdks81a, vdks81b, vdks82a},
an extensive sample of edge-on galaxies has been studied by \citet{dg98},
and has been re-analysed by \citet{kvdkdg02}. The scale parameters
correlate well with the rotation velocity of the galaxy, e.g. for
the scaleheight
\begin{equation}
h_{\rm z} = (0.45 \pm 0.05)\, (V_{\rm rot}/100\ {\rm km\  s}^{-1}) -
(0.14 \pm 0.07)\ {\rm kpc}
\end{equation}
with a scatter of 0.21 kpc. This could be useful to find a statistical
estimate of the thickness of disks in galaxies that are not seen edge-on.
The flattest galaxies (largest ratio of $h$ and $h_{\rm z}$) appear to be
those with late Hubble type, small rotation velocity and faint (face-on)
surface brightness. 

\section{Stellar kinematics}

At the basis of the analysis of the vertical dynamics of a stellar disk
we have the Poisson equation for the case of axial symmetry and at low $z$
\citep[e.g.][]{oort65}
\begin{equation}
{\D K_{\rm R} \over \D R} + {K_{\rm R} \over R} + {\D K_{\rm z} \over \D z} \approx 
2(A - B)(A + B) + {\D K_{\rm z} \over \D z} = 
-4\pi G \rho (R,z)
\label{eqn:Poisson}
\end{equation}
For a flat rotation curve $A = -B$ and $2(A-B)(A+B)=0$, 
so the plane-parallel case becomes an excellent approximation \citep{vdkf86} .
The equation of hydrostatic equilibrium 
\begin{equation}
\sigma_{\rm z}(R) = \sqrt{c \pi G \Sigma (R) h_{\rm z}},
\label{eqn:hydrostat}
\end{equation}
relates the vertical velocity dispersion 
$\sigma_{\rm z}(R)$ of the old stars to the surface density $\Sigma$ 
so that if the mass-to-light ratio $M/L$ is constant with radius, the exponential
radial surface brightness distribution 
implies that $\sigma_{\rm z}(R)$ should decline as the squareroot of 
$\Sigma$ or also as an exponential with radius,
but with an e-folding of twice the scalelength $h$ (the
constant $c$ in eqn.~(\ref{eqn:hydrostat}) varies between 3/2 for an exponential with
$n=\infty$ in eqn.~(\ref{eqn:vdksdisk}) to 2 for an isothermal distribution with $n=1$).
This was first tested in face-on spirals by \citet{vdkf84, vdkf86}, where the prediction
was verified in detail in NGC 5347; in fact, the e-folding of
$\sigma_{\rm z}$ was 2.4$\pm$0.6 photometric
scalelengths. Many studies have since confirmed this decrease of 
$\sigma_{\rm z}$ with radius \citep[e.g.][and
references therein]{bot93,kvdkf04, kvdkf05}. 

There are two recent developments that have a very strong impact on this
issue. The first is the use of integral field units that enable a more
complete sampling of the disks. The so-called {\it Disk Mass Project}
\citep{vbsaw07,wbvas08} aims at a mapping of the stellar vertical velocity
dispersion in this manner
in about 40 face-on spiral galaxies. As above this will provide a kinematic
measurement of the mass surface density of stellar disks. Not many results
have appeared in the literature yet, but recent conference presentations
show that the {\it `kinematics follows the light'}, i.e. the velocity dispersions
drop off according to the manner described above with constant $M/L$. The actual values
indicate relatively low mass-to-light ratios that are well
below those required for maximum disk fits (see below).

This result is also obtained in the other recent development, which is the
use of planetary nebulae as test particles in disks
\citep{hc09} of five face-on spirals. This method allows
the velocity dispersion of these representative stars of the old disk population
to be measured out to large radii. In general the findings are the same:
except for one system, the $M/L$ is constant out to about three radial
scalelengths of the exponential disks. Outside that radius the velocity
dispersion stops declining and becomes flat with radius. Possible
explanations these authors put forward for this behavior include an increase
in the disk mass-to-light ratio, an increase in the importance of the thick
disk, and heating of the thin disk by halo substructure. They also find that
the disks of early type spirals have higher values of $M/L$ and are closer to
maximum disk than later-type spirals.

There is certainly support from stellar dynamics that
in general there are no substantial gradients in mass-to-light ratios in disks.
The rather low $M/L$ values that are obtained currently do not require large
amounts of material unaccounted for, as was found originally by \citet{kap22}
and \citet{oort32}.

The stellar velocity dispersions {\it in} the plane are more complicated
to determine from observations. The
radial and tangential ones are not independent, but governed by the
local Oort constants: $\sigma_{\theta}/\sigma_{\rm R} = \sqrt{ -B/(A - B)}$.
This results from the axis ratio of the epicyclic motion that describes stellar
orbits deviating little from circular. The
frequency in the epicycle is $\kappa = 2\sqrt{-B( A - B)}$ and its axis
ratio $\sqrt{-B/(A - B)}$ \citep{oort65}.
For a flat rotation curve $A=-B$, so $\sigma_{\theta}/\sigma_{\rm R}$
(the `axis ratio of tghe velocity ellipsoid') is 0.71 and 
$\kappa = \sqrt{2}V_{\rm rot}/R$.

In highly inclined or edge-on systems the dispersions can be meaured
both from the line profiles and the asymmetric drift equation
\begin{equation}
V^2_{\rm rot} - V^2_{\theta} = \sigma^2_{\rm R} \left\{ {R \over h} -
R {\D \over {\D R}} \ln (\sigma_{\rm R}) - \left[ 1 - { B \over {B - A}}
\right] \right\},
\end{equation}
where the circular velocity $V_{\rm rot}$ can be measured with sufficient
accuracy from the gas (optical emission lines or \HI\ observations), which
have velocity dispersions of order 10 km/s or less and therefore
very little asymmetric drift ($V_{\rm rot} - V_{\theta}$).

The radial dispersion
plays an important role in the \citet{too64} $Q$-criterion for local
stability in galactic disks
\begin{equation}
Q = { {\sigma_{\rm R} \kappa} \over {3.36 G \Sigma}}
\end{equation}
with $\Sigma$ the local mass surface density. On
small scales local stability results from a Jeans-type stability, where the
tendency to collapse under gravity is balanced by the kinetic energy in random
motions, but only up to a certain (Jeans) scale. On large scales,
shear as a result of galactic
differential rotation provides stability. In the Toomre $Q$-criterion
the smallest scale for this is just equal to the Jeans scale, so that
that local stabiltiy exists on {\it all} scales. According to \citet{too64},
local stability requires $Q \gt 1$. Numerical simulations suggest that galaxy
disks have $Q =$ 1.5--2.5 and are on the verge
of instability \citep{hohl71,sc84,as86,mmb97,bot03}.

The first study where an attempt was made to measure these velocity
dispersions was by \citet{vdkf86} on the highly inclined
galaxy NGC 7184. They fitted their data using two different assumptions
for the radial dependence of the radial velocity dispersion, one being
that the axis ratio of the velocity ellisoid is the same everywhere, and
the other that Toomre $Q$ is constant with radius. Over the range from the
center of one or two scalelengths the assumptions work out to similar
variations \citep[see][page 196]{gkvdk90}.

\begin{figure}[t]
\begin{center}
\sidecaption[t]
\includegraphics[width=65mm]{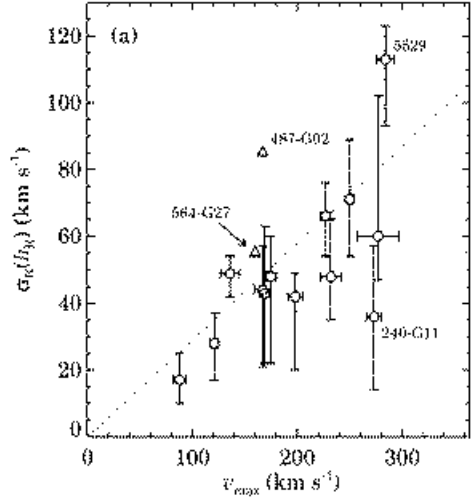}
        \caption{Stellar disk velocity dispersion, measured at one scalelength
in edge-on galaxies  versus the
        maximum rotational velocity. The gray lines indicate the relation
        $\sigma_{\rm R}(h) =$ (0.29$\pm$0.10) $V_{\rm rot}$
        \citep{bot93}. \citep[From][]{kvdkf05}}
\label{fig:7_bottema2}
\end{center}
\end{figure}

More extensive observations on a sample of 12 galaxies
\citep[including the Milky Way Galaxy from][]{lf89} by \citet{bot93}
resulted in the discovery of a relation between a fiducial value of the
velocity dispersion (either the vertical one measured at or
extrapolated to the center or the radial velocity dispersion
at one schalelength) and the integrated luminosity or
the rotation velocity (equivalent through the
Tully-Fisher relation). This has been confirmed (see fig.~\ref{fig:7_bottema2})
by \citet{kvdk05} and \citet{kvdkf05}:
\begin{equation}
\sigma_{{\rm z}|0} = \sigma_{\rm R|1h} = (0.29 \pm 0.10) V_{\rm rot}.
\label{eqn:bottema}
\end{equation}
It actually extends to small dwarf galaxies, e.g. 19 km sec$^{-1}$ in UGC 4325
\citep[][chapter 7]{swat99}.
Interestingly, the scatter in this relation is not random. 
Galaxies below the relation (with lower velocity dispersions) 
have higher flattening, lower central surface brightness or dynamical
mass ($4 h V^2_{\rm rot}/G$) to disk luminosity ratio.

The linear $\sigma-V_{\rm rot}$ relation follows from straightforward
arguments (\citealt{gkvdk90, bot93, vdkdg99}).
We evaluate now properties at one radial scalelength ($R = 1 h$). 
For a flat rotation curve and eliminating $h$
using a Tully-Fisher relation $L_{\rm disk}
\propto \mu _{\circ} h^{2} \propto V_{\rm rot}^{4}$ results in
\begin{equation}
\sigma _{\rm R} \propto Q \left( M \over L \right)_{\rm disk} \mu
_{0}^{1/2} V_{\rm rot}.
\label{eqn:botrel}
\end{equation}
This shows that when $Q$ and $M/L$ are constant among galaxies, the
Bottema relation itself results and galaxy disks
with lower (face-on) central surface brightness $\mu _{\circ}$ have
lower stellar velocity dispersions than the mean.

\section{Mass-to-light ratios and `maximum disk'
}
\begin{figure}[t]
\begin{center}
\sidecaption[t]
\includegraphics[width=65mm]{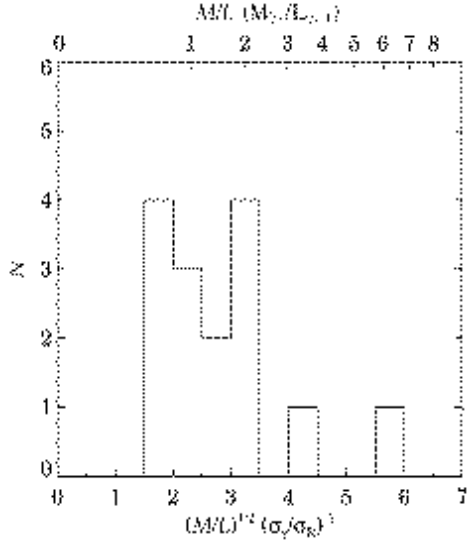}
   \caption{Histogram of the product $\sqrt{M/L}\
        (\sigma_{\rm z}/\sigma_{\rm R})^{-1}$ from stellar kinematics
in edge-on galaxies. Except for two outliers
        the distribution of $\sqrt{M/L}\ (\sigma_{\rm z}/\sigma_{\rm
        R})^{-1}$ is rather narrow. The outliers are ESO 487-G02 and
564-G27; data for these galaxies are less complete than for the other
ones. Along the top we show the values of
        $M/L$ implied by $\sigma_{\rm z}/\sigma_{\rm R}=0.6$.
\citep[From][]{kvdkf05}}
\label{fig:7_histogram}
\end{center}
\end{figure}

For a disk that is exponential in both the radial and vertical direction
\citep[cf.][]{vdk88}:
\begin{equation}
\sigma_{\rm z}(R, z) = \sqrt{\pi G h_{\rm z} (2 -
e^{-z/h_{\rm z}}) (M/L) \mu_{0}}\ e^{-R/2 h},
\label{eqn:sigma_z}
\end{equation}
Assuming a constant axis ratio of the velocity ellipsoid
$\sigma_{\rm z}/\sigma_{\rm R}$, we find
\begin{equation}
\sigma_{\rm R}(R,z) = \sqrt{\pi G h_{\rm z} (2 -
e^{-z/h_{\rm z}}) (M/L) \mu_{0}}\
\left( {{\sigma_{\rm z}} \over {\sigma_{\rm R}}} \right)^{-1}\
e^{-R/2 h}.
\label{eqn:sigma_R}
\end{equation}
The distribution of the products $\sqrt{M/L}\ (\sigma_{\rm
z}/\sigma_{\rm R})^{-1}$ in the \citet{kvdkf05} sample is
shown in fig.~\ref{fig:7_histogram}. Thirteen of the fifteen disks
have 1.8 $\simlt \sqrt{M/L}\ (\sigma_{\rm z}/\sigma_{\rm R})^{-1}
\simlt$ 3.3. The values of the outliers may have been overestimated
\citep[see][]{kvdkf05}. Excluding these, the
average is $\left<\right.\sqrt{M/L}\ (\sigma_{\rm z}/\sigma_{\rm
R})^{-1}\left.\right> =$ 2.5$\pm$0.2 with a $1\sigma$ scatter of 0.6.
The near constancy of the product can be used with $M/L$
based on stellar population synthesis models to estimate the axis
ratio of the velocity ellipsoid. Conversely, the upper scale of
Fig.~\ref{fig:7_histogram} indicates that a typical ($I$-band) $M/L$ 
of a galactic stellar disk is of order unity and varies for the
majority systems between 0.5 and 2.

It is possible to relate the axis ratio of the velocity ellipsoid to the
flattening of the stellar disk $h/h_{\rm z}$ \citep{vdkdg99}. In the
radial direction the velocity dispersion is related to the epicyclic
frequency through the Toomre parameter $Q$ for local stability. 
The Tully-Fisher relation then relates
this to the integrated magnitude and hence to the disk scalelength. 
In the vertical direction the scaleheight and
the velocity dispersion relate through hydrostatic equilibrium.

Eqn.~(\ref{eqn:botrel}) shows that when $Q$ and $M/L$ are constant among
galaxies, disks with lower (face-on) central surface brightness
$\mu _{\circ}$ have
lower stellar velocity dispersions.
Combining eqn.~(\ref{eqn:botrel}) with eqn.~(\ref{eqn:hydrostat}) for hydrodynamic
equilibrium  and using eqn.~(\ref{eqn:bottema}) gives \citep{kvdkf05, vdkdg99}
\begin{equation}
{h \over h_{\rm z}} \propto Q \left( {\sigma _{\rm R} \over \sigma
_{\rm z}} \right) \sigma _{\rm z}^{-1} V_{\rm rot}
\propto Q \left( {\sigma _{\rm R} \over \sigma _{\rm z}} \right).
\label{eqn:B1}
\end{equation}
The observed constancy of $\sqrt{M/L}\ (\sigma_{\rm z}/\sigma_{\rm R})^{-1}$
implies that the flattening of the disk $h/h_{\rm z}$ is proportional
to $Q\sqrt{M/L}$.

\begin{figure}[t]
\begin{center}
\sidecaption[t]
\includegraphics[width=70mm]{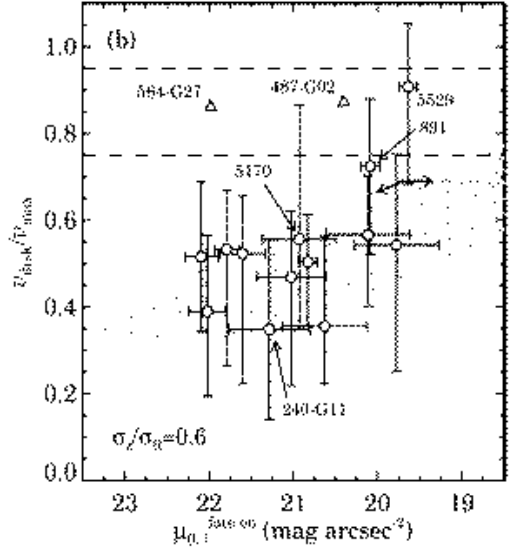}
\caption{The contribution of the disk to the amplitude
of the rotation curve $V_{\rm disk}/V_{\rm rot}$.
for a sample of 15 edge-on galaxies as a function
of the rotation velocity itself. The horizontal dashed lines are the limits
of $0.85 \pm 0.10$ from \citet{sac97}, which would indicate
maximal disks. The axis ratio of the velocity
ellipsoid is assumed to be 0.6. The grey lines correspond to collapse
models of \citet{dss97}. The two without error bars are the same
galaxies as the outliers in fig.~\ref{fig:7_histogram}.
\citep[From][]{kvdkf05}}
\label{fig:7_maximumdisk1}
\end{center}
\end{figure}

The mass-to-light ratio is a crucial measure of the contribution
of the disk to the rotation curve and the relative importance of disk
and dark halo mass in a galaxy. In the `maximum disk hypothesis' 
the disk contribution is optimized 
such that the amplitude of the disk-alone rotation curve is as large as the
observations allow. In a maximal disk, the ratio between the disk-alone rotation 
curve and the observed one will be a bit lower than unity to allow a bulge contribution
and let dark halos have a low density core. A working definition is
$V_{\rm disk}/V_{\rm rot}=$ 0.85$\pm$0.10 \citep{sac97}. 

For an exponential disk, the ratio of the peak rotation velocity of
the disk to the maximum rotation velocity of the galaxy
($V_{\rm disk}/V_{\rm rot}$) is
\begin{equation}
\frac{V_{\rm disk}}{V_{\rm rot}} = \frac{0.880\ (\pi G\,
\Sigma_{0}\, h)^{1/2}}{V_{\rm rot}}.
\label{eqn:7_freeman}
\end{equation}
Using eqn.~(\ref{eqn:hydrostat}) and eqn.~(\ref{eqn:bottema}) this can be
rewritten as
\begin{equation}
\frac{V_{\rm disk}}{V_{\rm rot}} = (0.21 \pm 0.08) \sqrt{h \over {h_{\rm
z}}}.
\end{equation}
So we can estimate the disk contribution to the rotation curve from
a statistical value for the flattening
\citep[see also][]{bot93, bot97, vdk02}. For the sample of \citet{kvdkdg02} this
then results in  $V_{\rm disk}/V_{\rm rot}=$ 0.57$\pm$0.22 (rms scatter).
In the dynamical
analysis of \citet{kvdkf05}, the ratio $V_{\rm disk}/V_{\rm rot}$
is known up to a factor $\sigma_{\rm z}/\sigma_{\rm R}$ and
distance-independent. For $\sigma_{\rm z}/\sigma_{\rm R}=0.6$, 
 $v_{\rm disk}/v_{\rm rot}=$ 0.53$\pm$0.04,
with a $1\sigma$ scatter of 0.15. Both estimates agree well. Thus, at least for
this sample, the average spiral has a submaximal disk. 

\begin{figure}[t]
\begin{center}
\includegraphics[width=50mm]{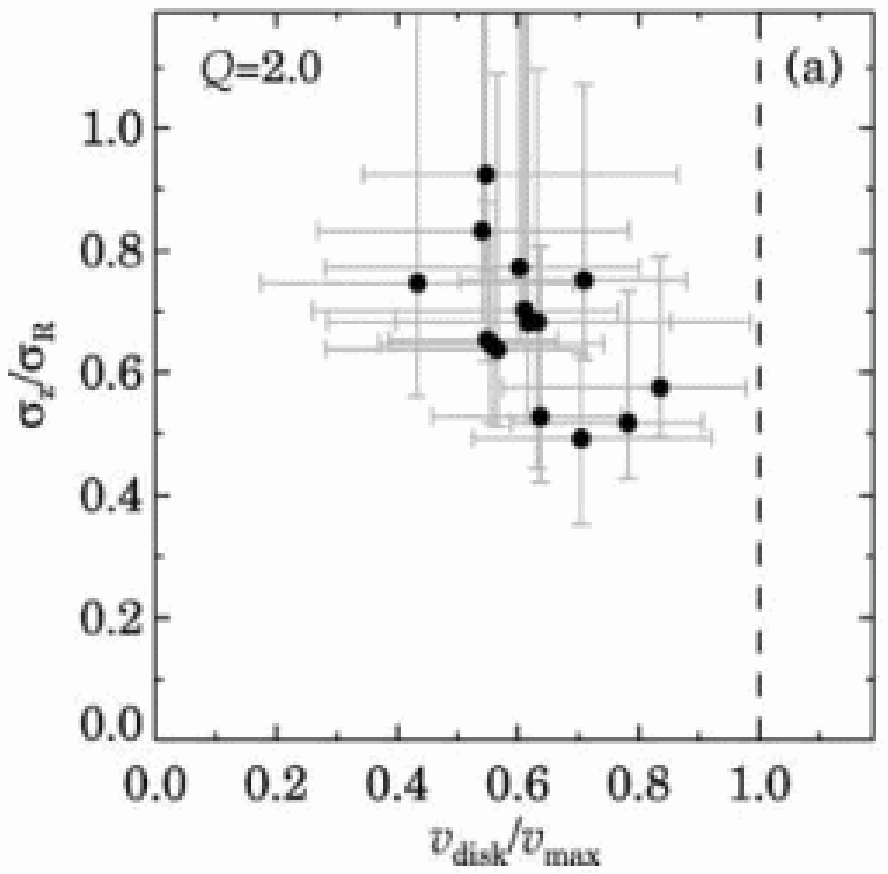}
\includegraphics[width=50mm]{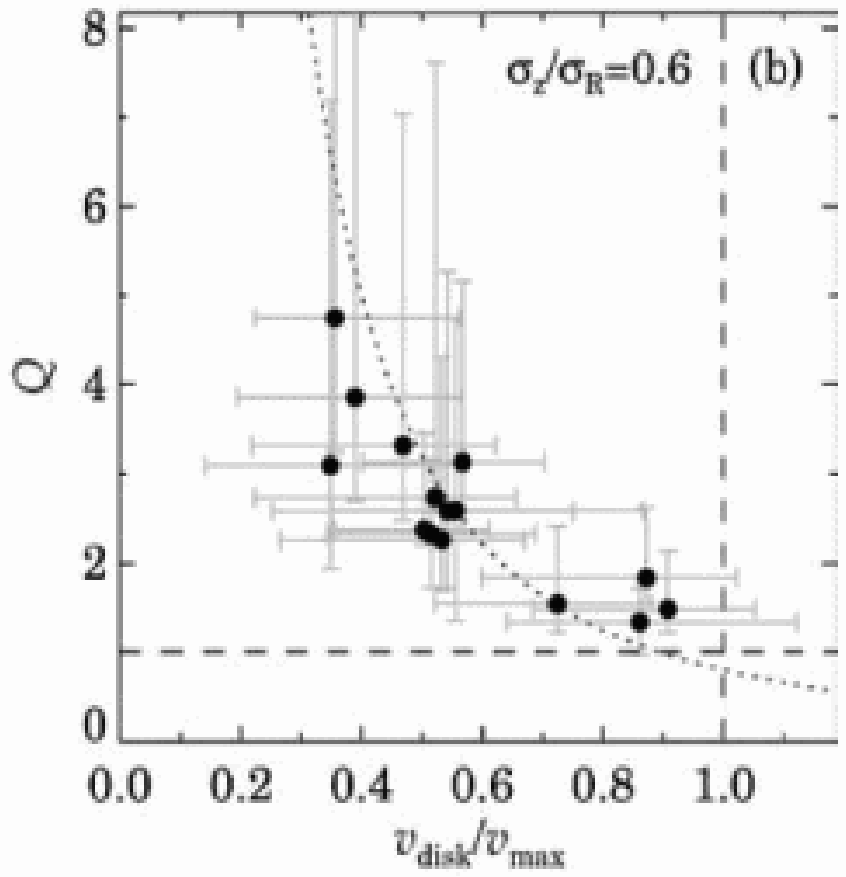}
        \caption{Stellar dynamics parameters for edge-on galaxies.
(a) The axis ratio of the velocity ellipsoid as
a function of $V_{\rm disk}/V_{\rm rot}$ for $Q$ = 2.0. (b)
$V_{\rm disk}/V_{\rm rot}$ as a function of $Q$ for an assumed axis ratio of
the velocity ellipsoid of 0.6.  \citep[From][]{kvdkf05}}
\label{fig:7_maximumdisk2}
\end{center}
\end{figure}

The values obtained for individual galaxies are illustrated in
figs.~\ref{fig:7_maximumdisk1} and \ref{fig:7_maximumdisk2}.
Most galaxies are not `maximum-disk'. The
ones that may be maximum disk have a high surface density
according to fig.~\ref{fig:7_histogram}. From the panels we also note
that disk that are maximal appear to have more anisotropic velocity
distributions or are less stable according to Toomre $Q$,

Originaly,
\citet{vdkf86} used the property $Y$ defined as a criterion for global stability
\citep{efn82}: $Y = V\subr{rot}\sqrt{h/G M\subr{disk}} \gsim 1.1$ for stability.
This is --up to a factor 0.62 for an infinitessimally thin exponential disk--
the reciprocal of the ratio $V_{\rm disk}/V_{\rm rot}$. Then they 
found $Y = 1.0 \pm 0.3$ \citep{vdkf86}, which corresponds to $V_{\rm disk}/V_{\rm rot} =
0.6 \pm 0.2$. It is remarkable --and probably significant-- that all values 
obtained for $V_{\rm disk}/V_{\rm rot}$ are close to 0.6.

\section{Flaring of HI layers and disk masses}

The thickness of the gas layer in a disk galaxy can be used to measure the surface
density of the disk. Assume the
density distribution of the exponential, locally isothermal disk (that in 
eqn.~(\ref{eqn:vdksdisk}) with $n=1$).
If the \HI\ velocity dispersion
$\langle V\subr{z}^{2} \rangle \subr{HI}^{1/2}$ is independent of radius --as e.g.
in the face-on spiral NGC 628, \citep{svdk84}-- {\it and} isotropic,
and if the stars dominate the gravitational field, the \HI\ layer  has a
full width at half maximum (to \lsim 3\%) of
\begin{equation}
W\subr{HI} = 1.7 \langle V\subr{z}^{2} \rangle \subr{HI}^{1/2} \left[
{\pi G(M/L) \mu\subr{\circ } \O z\subr{\circ }} \right] ^{-1/2} \exp (R/2h).
\label{eq:flaring}
\end{equation}
So the \HI\ layer increases exponentially in thickness with an e-folding $2h$.
This has first been derived and applied to \HI\ observations of NGC 891 by
\citet{vdk81}.
One has to be careful to distinguish signatures for flaring from those of
residual inclination away from exactly edge-on. Such studies can determine
whether galaxies have in general {\it maximum disks} or not.
The modelling, using photometry from \citet{vdks81b}, indicated
that $V\subr{rot,disk}$ of NGC 891 is $\sim 140$ km/s.
The observed value is $225 \pm 10$ km/s, so the ratio in eqn.(1)
is $\sim 0.6$, and NGC 891 is clearly sub-maximal.

For our Galaxy my preferred values are $V\subr{rot,disk} \sim
 155 \pm 30$ and $V\subr{rot,obs} \sim 225 \pm 10$ km/s, so that
the ratio is $0.69 \pm 0.14$ and the Milky Way also is sub-maximal.
In other systems similar results were found; e.g. in NGC 4244 \citet{olling96b} deduced
form the flaring a disk-alone rotation of 40 to 80\%\ of that
of observed rotation. Actually, the flaring of the \HI\ layer in NGC 4244
was used by \citet{olling96a} to infer that the dark matter is highly
flattened (but see \citet{om00}, who found for the Galaxy halo closer to
spherical). 

In a recent study by \citet{obfvdk10} on the superthin edge-on galaxy UGC 7321,
the rotation curve was decomposed using constraints from the thickness and
flaring of the \HI\ layer. This study was aimed at a determination of the shape
of the dark matter halo, which was found to be close to spherical. The disk ($I$-band) 
$M/L$ was found to be only about 0.2 and the galaxy is very far from maximum disk. 
This $M/L$ is even somewhat lower than the range 0.5 to 2 indicated above. In fact, 
\citep{bmj10} also concluded that the dark matter dominates the gravitational
field everwhere in the disk of this galaxy. 

Work on de compositions of rotation curves and analysis involving observations
of the `baryonic' Tuly-Fisher relation has progressed as well. In particular,
\citet{mg05} has analysed a sample of galaxies with extended \HI\ rotation
curves and finds that high surface brightness galaxies are closer to
maximum disk than low surface brightness ones. This may be the same
effect as described above in terms of surface density. On the other hand,
\citet{wsw01} find from detailed fluid dynamical gasflows in the barred
galaxy NGC 4123 that this system must be close to maximum-disk. So, almost
certainly some (massive) disks {\it are} maximal.

Finally I note the following clever
argument of \citet{cr99} that makes use of the {\it scatter} in the Tully-Fisher
relation. The amplitude of the rotation curve of the self-gravitating
exponential disk is
\begin{equation}
V_{\rm max} \propto \sqrt{M_{\rm disk}/h}.
\end{equation}
For fixed disk-mass $M_{\rm disk}$ we then get from differentiation
\begin{equation}
{{\D \log V_{\rm max}} \over {\D \log h}} = -0.5.
\end{equation}
So at a given absolute magnitude (or mass) a lower scalelength
disks should have a higher rotation velocity. If all galaxies were
maximum disk this then should be visible in the scatter of the
Tully-Fisher relation. This is {\it not} observed and the estimate
is that on average $V_{\rm disk} \sim 0.6 V_{\rm total}$ and galaxies
in general do not have maximal disks.

Recent reviews of disk masses in galaxies are by \citet{vdk09} and
\citet{mg09} at the Kingston symposium. In summary: the
overall state of affairs concerning the maximum disk hypothesis appears
to be that in general galaxy disk are {\it not} maximal, except possibly the
ones with the highest surface brightness and surface density.

\section{Flatness and truncations in stellar disks}

It is important to realise that stellar disks are often
remarkably flat. This can be studied in edge-on systems by
determining the centroid of the light distribution 
in the direction perpendicular to the major
axis at various galactocentric distances \citep[e.g.][chapter 5]{ssbf90,
fpbmss91, dg97}. Apart from some minor warps in the
outer parts of the stellar disks, in the inner parts the systematic deviatons are
very small.

\begin{figure}[t]
\begin{center}
\includegraphics[width=115mm]{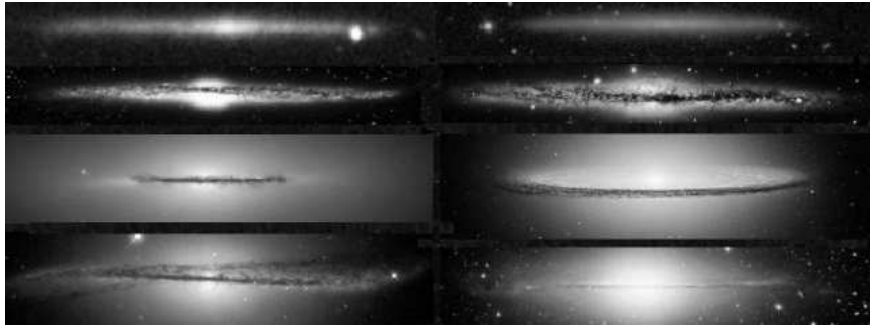}
        \caption{Selected images of edge-on disks and dust lanes
from various public Web-galleries.
Top: `Superthin' galaxies  IC 5249 \citep[from the
Sloan Digital Sky Survey,][]{vdkjvkf01}
and UGC 7321 (cosmo.nyu.edu/hogg/rc3/UGC\_732\_irg\_hard.jpg);
second row: NGC 4565
(www.cfht.hawaii.edu/HawaiianStarlight/AIOM/English/2004/Images/Nov-Image2003-CFHT-Coelum.jpg)
and NGC 891 (www.cfht.hawaii.edu/HawaiianStarlight/Posters/
NGC891-CFHT-Cuillandre-Coelum-1999.jpg);
third row: NGC 5866 (heritage.stsci.edu/ 2006/24/big.html) and
M104 (heritage.stsci.\-edu/2003/28/big.html);
bottom row: ESO 510-G013 (heritage.stsci.edu/2001/23/big.html) and
NGC 7814
(www.cfht.hawaii. edu/HawaiianStarlight/English/Poster50x70-NGC7814.html).
}
\label{fig:edge-ons}
\end{center}
\end{figure}

We may also look at the flatness of the layers of the ISM within them,
such as dustlanes. In fig.~\ref{fig:edge-ons} I collected some images of
edge-on disk galaxies. At the top are two `super-thin' galaxies;
the disks are straight lines to within a few percent. The same holds for
the dustlanes in NGC 4565 (allow for the curvature due to the imperfect
edge-on nature) and  NGC 891. In the third row the peculiar structure of NGC 5866
has no measuable deviation from a straight line, while for the Sombrero
Nebula the outline of the distlane fits very accurately to an ellipse.
In the bottom row, NGC 7814 (right) is straight again to wthin a few
percent, but NGC 5866 is an example of a galaxy with a large warp in the
dust layer.

The \HI\ kinematics provide probably the strongest indications for
flatness. In three almost completely face-on spirals (NGC 3938, 628 and
1058), \citet{vdksh82, vdksh84} and \citet{svdk84} found that the residual velocity 
field after subtraction of that of the rotation field 
has an r.m.s. value of only 3 - 4 km/s (or a few pc per Myr)
without any systematic pattern.
A vertical oscillation with a similar period as that for stars in the Solar Neighborhood
($10^7$ years) or even of that of rotation around the Galactic Center
($10^8$ years) would correspond to a vertical amplitude ten to a hundred pc.
The absence of such residual patterns shows that the \HI\
layers and the stellar disks must be
extraordinarily flat, except maybe in their outer regions or when they have recently been in
interaction.

Recently, \citet{mu08a, mu08b} have found evidence for a pattern of corrugation
in the disk of the edge-on galaxy IC 2233 with an amplitude up to 250 pc, especially in 
\HI\ and young stars. IC 2233
is a rather small galaxy (radius 7 or 8 kpc and rotation velocity about 100 km/s) 
unlike the ones discussed in the previous paragraph.

I have indicated above that the flattening of the stellar disk
$h_{\rm z}/h$ is smallest for systems of late Hubble type,
small rotation velocity and faint (face-on) surface brightness. It is of
interest then to look more closely at systems at this extreme end of
the range of flattening: `superthin' galaxies. A prime example is the
galaxy UGC 7321, studied extensively by Lynn Matthews and collaborators 
\citep[][and references therein]{bmj10}.
The picture that appears is that this is
a very low surface brightness galaxy (the face-on $B$-band 
central surface brightness is $\sim$23.4 mag arcsec$^{-2}$) and a
scalelength of about 2 kpc, but a projected vertical scaleheight of only
150 pc. It appears to have vertical structure since there is a color
gradient (bluer near the central plane) and appears to consist of two
components. Its \HI\ is warped in the outer parts, starting at the edge of the light
distribution. 

Another good example of a superthin galaxy is IC 5249 \citep{byun98,
abc99, vdkjvkf01}. This also is a faint surface brightness galaxy with
presumably a small fraction of the mass in the luminous disk. However,
the disk scaleheight is not small (0.65 kpc). It has
a very long radial scalelength (17 kpc); the faint surface brightness
then causes only the parts close to the plane to be easily
visible against the background sky, while the long radial scalelength
assures this to happen over a large range of $R$. Therefore it appears
thin on the sky. The flattening $h_{\rm z}/h$  is 0.09 (versus 0.07 for
UGC 7321). The stellar
velocity dispersions are similar to those in the Solar Neighborhood;
disk heating must have proceeded at a pace comparable to that in the
Galaxy.

The flattest galaxies on the sky have indeed
very small values of $h_{\rm z}/h$. However, these two examples show
that superthin galaxies share at least
the properties of late type, faint face-on surface brightness and
small amounts of luminous disk mass compared to that in the dark halo.

\begin{figure}[t]
\begin{center}
\includegraphics[width=115mm]{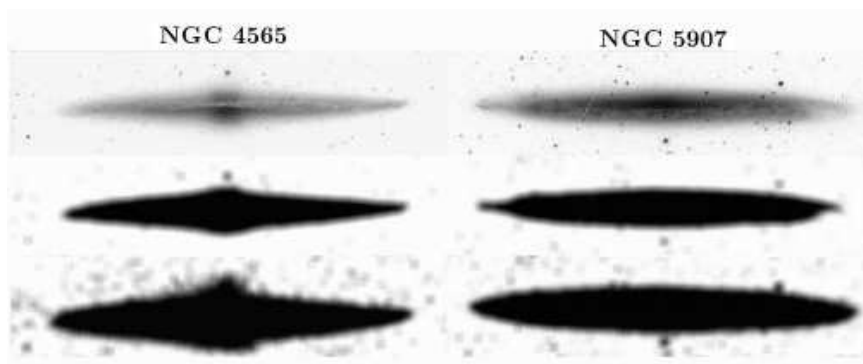}
\caption{NGC4565 and NGC 5907 at various light levels. These have been
produced from images of the Sloan Digital Sky Survey, which were clipped at
three different levels (top to bottom) and turned into two-bit
images adn subsequnetly smoothed
\citep[see][for an explanation of the details]{vdk07}. Note that the
disks grow significantly along the minor axes but not in radial extent.}
\label{fig:truncations}
\end{center}
\end{figure}

Truncations in stellar disks were first found in edge-on galaxies, where the
remarkable feature was noted that the radial extent did not grow
with deeper photographic exposures \citep{vdk79}. 
Prime examples of this phenomenon of truncations  are the large edge-on galaxies
NGC 4565 and NGC 5907 (see fig.~\ref{fig:truncations}).
The truncations appear very sharp, although of course not
infinitely so. Rather sharp outer profiles are actually obtained
after deprojecting near-IR observations of edge-on galaxies
\citep[e.g.][]{flo06}.

Various models have been proposed for the origin of truncations \citep[for a
discussion see][]{vdk07}. Truncations could be the current extent of
the disks while they are growing from the
inside out from accretion of external material.
This predicts larger age gradients across disks than are observed
\citep{dj96}. Another possibility is that
star formation is inhibited when the gas surface (or space?) density
falls below a certain threshold for local stability \citep{fe80,ken89,js04}.
The Goldreich--Lynden-Bell criterion for stability of gas layers
gives a poor prediction for the truncation radii \citep{vdks82a}.
Another problem for teh threshold hypothesis 
is that the rotation curves of some galaxies, e.g. NGC 5907
and NGC 4013 \citep{sc83,bot96}, show features
near the truncations that indicate that the {\it mass}
distributions are also truncated.
Schaye predicts an anti-correlation between $R_{\rm max}/h$ and $h$,
which is not observed. Models have been proposed \citep[e.g.][]{flo06}, 
in which a magnetic force breaks down as a result of star formation so that stars escape.
The evidence for sufficiently strong magnetic fields
needs strengthening.

Obviously, the truncation corresponds to the maximum
in the current specific angular momentum distribution of the disk,
which would correspond to that in the protogalaxy \citep{vdk87}
if the collapse occurs with detailed conservation of specific
angular momentum. As noted above, if this starts out as a  \citet{mes63}
sphere (i.e. uniform density and
angular rotation)   in the force field of a dark
halo with a flat rotation curve, a roughly exponential disk
results. This disk has then a truncation at about 4.5 scalelengths, so
this hypothesis provides at the same time an explanation for the exponential
nature of disk as well as for the occurence of the truncations.
This requires the absence of substantial redistribution of angular
momentum takes place.
Bars may play an important role in this, as suggested by \citet{dmcmwq06}
and \citet{erw07}.
In fact a range of possible agents in addition to bars, such as density waves,
heating and stripping of stars by bombardment of dark matter subhalos, has
been invoked \citep{djetal07}.

\begin{figure}[t]
\includegraphics[width=115mm]{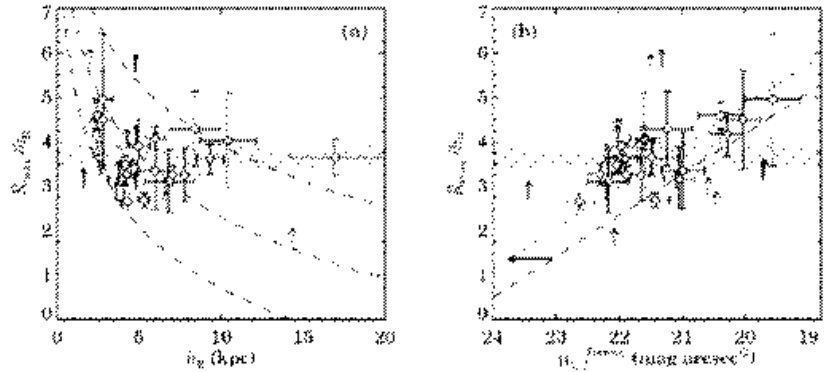}
\caption{Correlations of $R_{\rm max}/h$ with scalelength $h$ and
face-on central surface brightness $\mu _{\circ,{\rm fo}}$
for a sample of edge-on galaxies.
 The cross-hatched regions show the prediction from
a collapse model as in  \citet{vdk87} and \citet{dss97};
the dotted and dashed lines show
predictions from the star formation threshold model of \citet{js04} for
three different values of the disk mass \citep[from][see there for details]{kvdk04}}
\label{fig:truncs}
\end{figure}

\citet{kvdk04} derive correlations of the ratio of the truncation radius $R_{\rm max}$ 
and the disk scalelengths $h$ with $h$ itself and with the face-on central surface
brightness $\mu _{\circ,{\rm fo}}$ (fig.~\ref{fig:truncs}).
$R_{\rm max}/h$ does not depend strongly on $h$, but is somewhat less
than the 4.5 predicted from the collapse from a simple Mestel-sphere.
There is some correlation between $R_{\rm max}/h$ and $\mu _{\circ, {\rm fo}}$,
indicating approximate constant disk surface density at the truncations,
as possibly expected in the star-formation theshold model.
But this model predicts an anti-correlation between $R_{\rm max}/h$ and
$h$ \citep{js04}, which is not observed.
The maximum angular momentum hypothesis predicts that
$R_{\rm max}/h$ should not depend on $h$ or $\mu _{\circ, {\rm fo}}$ and
such a model therefore requires some redistribution of angular momentum 
or somewhat different initial conditions.

Due to line-of-sight integration, truncations will be more difficult to
detect in face-on galaxies. The expected surface brightness at 4 scalelengths is
about 26 B-mag arcsec$^{-2}$ or close to sky. In face-on galaxies like
NGC 628 \citep{svdk84,vdk88} an isophote map shows that
the outer contours have a much smaller spacing than the inner ones.
The usual analysis uses an inclination and major axis determined from
kinematics (if available, otherwise this is estimated from the
average shape of isophotes) and then determines an azimuthally averaged
radial surface brightness profile. But this will smooth out any truncation
if its radius is not exactly constant with azimuthal angle.
The effects are nicely illustrated in the study of NGC 5923
\citep[][their fig.~9]{pdla02}, which has isophotes in polar coordinates.
The irregular outline shows that some smoothing out will occur contrary
to observations in edge-on systems.

\citet{pt06} studied a sample of moderately inclined systems through ellipse-fitting of 
isophotes in SDSS data.
They distinguish three types of profiles:
{\it Type I}: no break;
{\it Type II}: downbending break;
{\it Type III}: upbending break.
\citet{pzpd07} have reported that the same types profiles occur among edge-on
systems; however, of their 11 systems there were only one for each of
the types I and III.

Various correlations have been reviewed in \citet{vdk09}. In general, the
edge-on and face-on samples agree in the distribution of $R_{\rm max}/h$;
however the fits in moderately inclined systems result in small values of the
scalelength compared to the edge-on sample. S. Peters, R. S. de Jong and I have 
re-analysed the Pohlen et al. data using two approaches: (1) mimick an edge-on view by 
collapsing the data onto the major axis and (2) calculate a radial profile using 
equivalent profiles. The luminosity profiles from ellipse fitting 
(Pohlen et al.) and that using equivalent profiles agree well,
in spite of the difference that the first assumes a position for the center and
the method with equivalent profiles does not. Often the `major-axis-collapse' method 
shows in Types I {\it and} II truncations when seen `edge-on'. So, there are truncations 
in the stellar disks but less symmetric than one might expect. Finally, Type III 
galaxies do not show 'edge-on truncations', but invariably evidence for interaction or 
other disturbances of the outer parts.
A prime example of a Type III profile is NGC 3310, which is a 
well-known case of a disturbed, probably merging galaxy \citep{vdk76, ks01}.

There is a good correlation between $R_{\rm max}$
and the rotation velocity \citep{vdk08}. On average a galaxy like our own
would have an $R_{\rm max}$ of 15 - 25 kpc (and a scalelength of 4 - 5 kpc).
Now look at NGC 300 , which has no truncation even at 10 scalelengths 
\citep{bh05}, so that $R_{\rm max} \gt$ 14.4 kpc. In spite of that it is not
outside the distribution observed in edge-on systems between $R_{\rm max}$
and $V_{\rm rot}$ (NGC 300 has $\sim 105$ km/s and this would give an $R_{\rm
max}$ of 8 - 15 kpc and an $h$ of 2 - 4 kpc ). So it has a unusually small $h$ 
for its $V_{\rm rot}$; not an unusual $R_{\rm max}$ for its rotation!
At least some of the Type I galaxies could have
disks with normal truncation radii, but large $R_{\rm max}/h$ and
small $h$ so that the truncations occur at much lower surface brightness.

I note, but cannot discuss in detail, that truncations in stellar
disks and warps of \HI\ layers are often associated, and refer to my discussion
in \citet{vdk07}.

\begin{acknowledgement}
Ken Freeman is an expert in many area's of astronomy, but he is in particular known 
for his research in that of disks of spiral galaxies, and I feel fortunate to have been
able to work with him on projects related to that. Many congratulations, Ken, and thanks
for all the years of friendship and stimulating collaboration. 
\end{acknowledgement}
%
%
%

%
%
%

\end{document}